\documentclass[a4paper,11pt]{article}
\pdfoutput=1 

\usepackage{jheppub} 

\usepackage[T1]{fontenc} 
\def\be{\begin{equation}}
\def\ee{\end{equation}}
\def\bea{\begin{eqnarray}}
\def\eea{\end{eqnarray}}

\def\p{\partial}

\def\>{\rangle} 
\def\<{\langle} 

\title{\boldmath Thermodynamics and holographic entanglement entropy
for spherical black holes in 5D Gauss-Bonnet gravity}

\author[1]{Yuan Sun,}
\author[2]{Hao Xu,}
\author[3]{Liu Zhao}

\affiliation[]{School of Physics, Nankai University,
Tianjin 300071, China\\}

\emailAdd{sunyuan14@mail.nankai.edu.cn}
\emailAdd{haoxu@mail.nankai.edu.cn}
\emailAdd{lzhao@nankai.edu.cn}

\abstract{The holographic entanglement entropy is studied numerically in (4+1)-dimensional
spherically symmetric Gauss-Bonnet AdS black hole spacetime with compact boundary.
On the bulk side the black hole spacetime undergoes a van der Waals-like phase transition
in the extended phase space, which is reviewed with emphasis on the behavior on
the temperature-entropy plane.
On the boundary, we calculated the regularized HEE of a disk region of different sizes.
We find strong numerical evidence for the failure of equal area law for isobaric curves
on the temperature-HEE plane and for the correctness of first law of entanglement entropy,
and briefly give an explanation for why the latter may serve as a reason for the former,
i.e. the failure of equal area law on the temperature-HEE plane.}

\begin{document}
\maketitle
\flushbottom

\section{Introduction}
\label{sec:intro}

The gauge-gravity duality \cite{Maldacena:1997re} is a promising way to make a connection
between general relativity and quantum field theory. According to AdS/CFT correspondence,
the black holes in AdS spacetime are dual to strongly-coupled large $N$ gauge theories at
finite temperature. The thermodynamics of black hole predicts phase transition in AdS
spacetime, such as the Hawking-Page phase transition in Schwarzschild-AdS spacetime \cite{HawkingPage:1983}, which can be explained as the gravitational dual of QCD
confinement/deconfinement transition \cite{Witten:1998zw,Witten:1998qj}. Black
thermodynamics is also studied for charged black holes, and a
first order phase transition was found in Reissner-Nordstr\"om AdS (RN AdS) spacetime \cite{ChamblinEtal:1999a,ChamblinEtal:1999b}.

Recently, this analogy is extended to more general cases. By identifying the negative
cosmological constant as an effective pressure $P=-\frac{\Lambda}{8\pi}$,
the thermodynamics of black hole can be established on the extended phase space \cite{KastorEtal:2009,D.Kubiznak}. The physical meaning of the thermodynamical volume
that is conjugate to the effective pressure $P$ remains
to be fully understood, but it is conjectured to satisfy the reverse isoperimetric
inequality \cite{CveticEtal:2011}. In this consideration, the black hole mass is taken
as the enthalpy $H$ rather than the internal energy. The extended phase space
thermodynamics has been investigated for many different spacetimes \cite{Poshteh:2013pba,Altamirano:2013ane,Altamirano:2014tva,Wei:2012ui,
Cai:2013qga,Zou:2014mha,Gunasekaran:2012dq,Zou:2013owa,Johnson:2014xza,Johnson:2014pwa,
Xu:2013zea,Xu:2014tja,Xu:2014kwa,Frassino:2014pha,Dolan:2014vba,Lee:2014tma,
Rajagopal:2014ewa,Frassino:2015oca,Lee:2015wua}, and in many cases
the extended phase space thermodynamic behavior is very
similar to van der Waals liquid-gas system.

Up to now the dual field theory interpretation of the van der Waals-like phase transition
remains unknown. However, progress has been made in this direction recently. In
Ref. \cite{Johnson:2013dka}, it was found that the holographic entanglement entropy (HEE)
as a function of temperature behaves qualitatively the same as black hole entropy in
the context of a charged black hole in AdS background with finite volume. In this case,
the HEE undergoes van der Waals-like phase transitions, and an inflexion point appears on
the temperature-HEE curve at the same critical temperature. More recently, the similarity
between the two kinds of entropies has been investigated further by considering
Maxwell's equal area law, which holds for black hole entropy, and seems to be still valid
on the HEE-temperature curve \cite{Nguyen:2015wfa}. The numerical results show that for
RN-AdS black holes this ``equal area law'' on the HEE-temperature curve holds up to an
accuracy of around 1\%, however, it fails for dyonic RN-AdS black holes. Therefore, to get
a better understanding of the field theory interpretation of the van der Waals-like phase
transitions, it is important to examine whether these ideas applies to other gravity models.
This connection has been extended to other cases \cite{Caceres:2015vsa,Zeng:2015tfj,
Zeng:2015wtt,Dey:2015ytd}, including the extended phase space. It seems that the HEE can
be a nice probe of the extended phase space.

Motivated by the above considerations and progresses, we extended the study of van der
Waals-like behavior for HEE to Gauss-Bonnet AdS black holes with a spherical horizon in
(4+1)-dimensions. The thermodynamics of this particular black hole spacetime has been
studied in the extended phase space in \cite{Cai:2013qga}. It was shown that for GB-AdS
black holes, the $P-V$ criticality and phase transition only occurs when the black hole
has a spherical horizon. When the charge of the black hole is turned off, only in
(4+1)-dimension the $P-V$ criticality and phase transition takes place.

The inclusion of Gauss-Bonnet term is a non-trivial generalization of Einstein gravity.
As a consequence, one must employ the HEE formula for general higher derivative gravity \cite{Hung:2011xb,Lewkowycz:2013nqa,Dong:2013qoa,Miao:2014nxa}. We will show that that
the equal area law on the temperature-HEE plane fails but a van de Waals-like behavior
on both the temperature-HEE and the temperature-black hole mass curves indeed holds.

The rest of the paper is organized as follows: in section 2, we review the black hole
thermodynamics for spherically symmetric GB-AdS black holes, and discuss the critical
behavior and Maxwell equal area law on the entropy-temperature plane. In section 3, we briefly
review the holographic entanglement entropy in Gauss-Bonnet gravity and present the HEE formula
for our setup. In section 4, The numerical results are presented, which include,
in particular, the numerical evidence for the failure of the equal area law on the
temperature-HEE plane and the correctness of the first law of entanglement entropy,
which has never been established before. By employing the linear relationship between
HEE and black hole mass, we give an explanation for why the equal area law fails on the
the temperature-HEE plane. In the final section, we present some concluding remarks.

\section{Thermodynamics for Gauss-Bonnet AdS Black holes}

In this section, we give a brief review of the thermodynamics of Gauss-Bonnet AdS Black holes.
The detailed calculation can be found in \cite{RGCai2002,Cai:2013qga,Xu:2015hba}. The action of
Gauss-Bonnet gravity in $(d+1)$ dimensions can be written as \cite{RGCai2002}
\be \label{action}
\mathcal{I}=\frac{1}{16\pi G}\int \mathrm{d}^{d+1}x\sqrt{-g}\,[R-2\Lambda+\alpha_{GB}\mathcal{L}_{GB}]
\ee
where $\alpha_{GB}$ is the Gauss-Bonnet coefficient and
\be
\mathcal{L}_{GB}=R^2-4R_{ab}R^{ab}+R_{abcd}R^{abcd}
\ee
is known as Gauss-Bonnet density. The Gauss-Bonnet coefficient can be
identified with the inverse string tension with positive value in string theory, so we shall only
consider the case $\alpha_{GB}>0$ in this paper. For later convenience, we reparametrize
the negative cosmological constant as $\Lambda=-d(d-1)/(2L^2)$ and
the Gauss-Bonnet coefficient as $\alpha=\alpha_{GB}(d-2)(d-3)$.

Gauss-Bonnet gravity admits pure AdS solution with Riemann tensor $R_{abcd}=-(g_{ca}g_{bd}-g_{cb}g_{ad})/\tilde{L}^2$ and the radius $\tilde{L}$ is given by
\be
\tilde{L}^2=\frac{2\alpha}{1-\sqrt{1-\frac{4\alpha}{L^2}}}.
\ee
It follows that the Gauss-Bonnet coefficient must satisfy $\alpha\leqslant L^2/4$ in
order that the pure AdS solution exists.
Besides, there exists another constraint by demanding the causality of dual field theory
\cite{Zeng:2013mca}
\begin{equation}
-\frac{(3d+2)(d-2)}{4(d+2)^2}L^2\leqslant\alpha
\leqslant\frac{(d-2)(d-3)(d^2-d+6)}{4(d^2-3d+6)^2}L^2.
\end{equation}

In this paper, we will be interested in the AdS black hole solutions
which takes the form \cite{Boulware,RGCai2002,Wiltshir,Cvetic,Kofinas:2006hr,Kastor:2011qp,
Cai:2013qga}
\be
\mathrm{d}s^2_{(d+1)}=-f_{(d+1)}(r)\mathrm{d}t^2+\frac{1}{f_{(d+1)}(r)}\mathrm{d}r^2
+r^2h_{ij}\mathrm{d}x^i\mathrm{d}x^j,
\ee
where
\be
f_{(d+1)}(r)=k+\frac{r^2}{2\alpha}\left(1\pm\sqrt{1+\frac{64\pi G\alpha M}{(d-1)\Sigma_kr^d}
-\frac{4\alpha}{L^2}}\right),
\ee
$h_{ij}$  is the metric on the $(d-1)$-dimensional hypersurface with constant curvature
$(d-1)(d-2)k$ and volume $\Sigma_k$ with $k=-1,0,1$, $M$ is black hole mass.
Among the two branches of solutions, only the ``$-$'' branch is ghost free, so we
shall only consider this case in this work.  Moreover, we shall restrict ourselves only to
the spherically symmetric case by taking $k=1$.

In AdS background, the black hole event horizon $r_+$ is the largest root of $f_{(d+1)}(r)$.
The enthalpy and temperature can be obtained by using two equations $f_{(d+1)}(r_+)=0$ and
$T=\frac{f_{(d+1)}'(r_+)}{4\pi}$, which yield
\begin{equation} \label{mass}
 H\equiv M=\frac{(d-1)\Sigma_1 r_+ ^{d-2}}{16\pi}\bigg(1+\frac{\alpha}{r_+^2}
 +\frac{16\pi P r_+^2}{d(d-1)}\bigg),
\end{equation}
\begin{equation}
T=\frac{16\pi P r_+ ^4/(d-1)+(d-2) r_+ ^2
+(d-4) \alpha}{4\pi r_+  (r_+ ^2+2 \alpha)},
\label{T1}
\end{equation}
where the pressure $P$ is defined as $P=-\frac{\Lambda}{8\pi}$.

The black hole entropy \cite{RGCai2002} and thermodynamic volume can also be easily calculated
\begin{equation}
S=\frac{\Sigma_1 r^{d-1}_+}{4}\bigg[1+\frac{2(d-1)\alpha }{(d-3)r_+^2}\bigg],
\label{S}
\end{equation}
\begin{equation}
V=\bigg(\frac{\partial H}{\partial P}\bigg)_{S}=\frac{\Sigma_1 r_+ ^{d}}{d}.
\label{vol}
\end{equation}
These quantities satisfy the first law of black hole thermodynamics in the extend phase
space \cite{Xu:2015hba}, i.e.
\be
\mathrm{d}H=T\mathrm{d}S+V\mathrm{d}P, 
\label{firstlaw}
\ee
or, in terms of Gibbs free energy $G(T,P)=H-TS$,
\be \label{f2}
\mathrm{d}G=V\mathrm{d}P-S\mathrm{d}T. 
\ee

As discussed in \cite{Cai:2013qga}, there exists some $P-V$ critical behavior and first-order
phase transition in this extended phase space in $d+1=5$ dimensions, which is similar to the van
der Waals liquid-gas system.  One can perform similar analysis on the $T-S$ plane by fixing  $P$
in eq.(\ref{f2}). The critical point can be determined by solving following two equations
\be\label{cri}
\bigg(\frac{\partial T}{\partial S}\bigg)_{P}=0,~~
\bigg(\frac{\partial^2 T}{\partial S^2}\bigg)_{P}=0.
\ee
For ease of the forthcoming numerical calculations, we set $\alpha=1,\Sigma_1=1$ in the rest of
the paper. Then the above equations give rise to the following critical pressure and critical
radius for the black hole,
\be
~~P_c=\frac{1}{48\pi},~~r_c=\sqrt{6}.
\ee
The corresponding critical temperature is then
\be
T_c=\frac{\sqrt{6}}{12\pi}.
\ee

\begin{figure}[htbp]
\centering
\includegraphics[width=\textwidth]{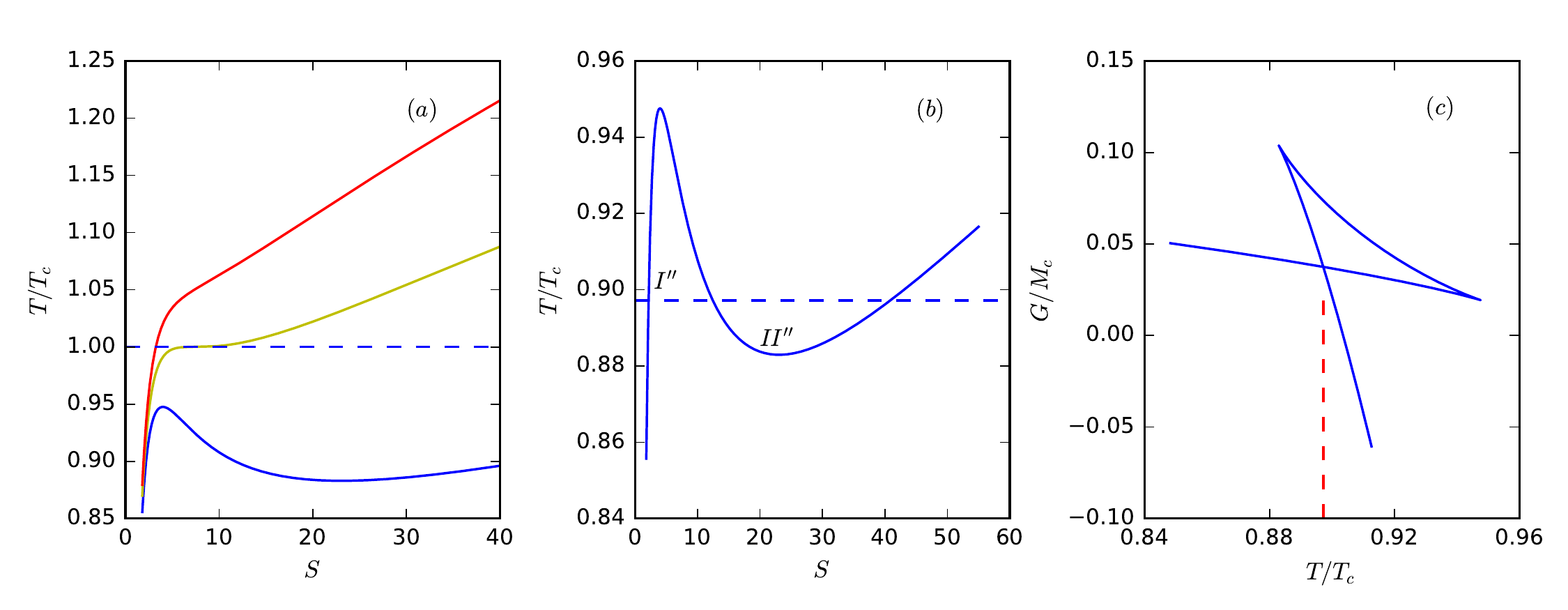}
\caption{(a) Isobaric curves on (4+1)-dimensional Gauss-Bonnet AdS black hole with different
pressure. From bottom to top the corresponding pressure are 0.7$P_c$(blue), $P_c$(orange),
$1.2P_c$(red). The dashed horizontal line(green) located at $T=T_c$. (b) Zoom in of the
$P=0.7P_c$(blue) curve in (a). The dashed horizontal line corresponds to $T=T^*$. (c) Gibbs
free energy along the isobaric curve in (b),  The dashed vertical line corresponds to  $T=T^*$.}
\label{fig:1}
\end{figure}

In Ref. \cite{Xu:2015hba}, the Maxwell's equal area law for Gauss-Bonnet AdS black holes is
studied on the $P-V$ plane. In the following we shall reconstruct it on the $T-S$ plane.
Fig.\ref{fig:1}$(a)$ gives the isobaric curves on the $T-S$ plane at pressures
$P=0.7P_c,P_c,1.2P_c$ are presented. It can be seen that, when $P$ is lower than the critical
pressure $P_c$, there may exist three black holes of different sizes at the same temperature.
However, the medium sized black hole is unstable since its heat capacity $C=T(\p S/\p T)$  is
negative. Naturally,  one wishes to know how Gibbs free energy varies along the isobaric curve.
As is shown in  Fig.\ref{fig:1}$(c)$, there is an intersection point at $T=T^*<T_c$ on the
isobaric Gibbs free energy versus temperature curve, which implies that at this temperature
the small black hole may jump into a large black hole. This is a first order phase transition
similar to the phase transition studied in the $P-V$ plane. The value of $T^*$ can be determined
by the Maxwell's equal area law since the Gibbs free energy remains unchanged during the phase
transition. By referring to the closed regions formed by the isobaric $T-S$ curve and the
isotherm $T=T^\ast$ as $\text{I}^{\prime\prime}$ and $\text{II}^{\prime\prime}$
(see Fig.\ref{fig:1}(b)), the equal area law can be expressed symbolically as
\be
\text{A}(\text{I}^{\prime\prime})=\text{A}(\text{II}^{\prime\prime}), \label{ar1}
\ee
where 
\[
\text{A}(\text{I}^{\prime\prime})\equiv \int_{S_1}^{S_3}\left|T(S)-T^\ast\right|\mathrm{d}S,
\quad
\text{A}(\text{I}^{\prime\prime})\equiv \int_{S_3}^{S_2}\left|T(S)-T^\ast\right|\mathrm{d}S,
\]
$S_1$, $S_2$ and $S_3$ are respectively the smallest, largest and the intermedium 
solution of the equation $T(S)=T^\ast$. Equivalently, we can also re-express \eqref{ar1}
as
\be
A_L\equiv T^*(S_2-S_1)=\int_{S_1}^{S_2}T(S)\mathrm{d}S\equiv A_R. \label{alr}
\ee
However, when calculating the relative disagreements between the areas of the two closed 
regions, one should avoid using this latter expression, because neither $A_L$ nor $A_R$ 
corresponds to the area of any of the closed regions.

When $P=P_c$, there is an inflection point on the isobaric curve,
and the area of both closed regions mentioned shrinks to zero. In this case, the size of the
three black holes at the same temperature becomes identical, and the phase transition
becomes continuous and is of the second order. If the pressure $P$ increases further so that
$P>P_c$ holds, $T$ becomes a monotonous function of $S$, and there can be only a single black
hole at each temperature, therefore the phase transition no longer occurs.

\section{HEE in Gauss-Bonnet AdS gravity}

Nowadays much attention has been focused on the research of entanglement entropy, which appears
in many fields of physics, such as {quantum field theory{\cite{Calabrese:2004eu,Casini:2009sr}}}, condensed matter
physics\cite{Amico:2007ag}, and quantum information\cite{Vedral}. For a quantum system with
density matrix $\rho$, the entanglement entropy of a subsystem $A$ is defined as
\be
S_A=-\text{Tr}\rho_A\ln\rho_A,
\ee
where $\rho_A$ is the reduced density matrix of $A$, which is defined by tracing over the
degrees of freedom in the complementary subsystem $\bar{A}$ of $A$, i.e.
\be
\rho_A=\text{Tr}_{\bar{A}}\rho.
\ee
In the framework of AdS/CFT, Ryu and Takayanagi conjectured that the entanglement entropy of
the dual field theory can be calculated holographically from the gravity side \cite{Ryu:2006bv,Ryu:2006ef} using the formula
\be \label{rt}
S_A=\frac{\text{Area} (\Sigma)}{4G},
\ee
where $\Sigma$ is the co-dimension 2 minimal surface whose boundary coincides with the
entangling surface between $A$ and $\bar{A}$, and $G$ is the Newton constant of the bulk theory.
This formula applies to Einstein gravity. This geometric description is reminiscent of the
Bekenstein-Hawking entropy for black holes since both are proportional the area of some
surfaces. Moreover, the similarity between the two kinds of entropies can go beyond Einstein
gravity\cite{Hung:2011xb}. For Lovelock gravity the holographical entanglement entropy is
calculated by minimizing a certain surface functional which is originally used to compute black
hole entropy in Lovelock gravity \cite{Jacobson:1993xs} (this surface functional is denoted as
$S_{\text{JM}}$ in \cite{Hung:2011xb}). The general formula for HEE in higher curvature gravity
is given in \cite{Dong:2013qoa,Miao:2014nxa}, which is based on the generalized gravitational
entropy introduced in \cite{Lewkowycz:2013nqa}.
For Gauss-Bonnet gravity the holographic entanglement entropy takes the form
\cite{Hung:2011xb,Dong:2013qoa}
\be\label{GBHEE}
S_{A}=\frac{1}{4 G}\int_\Sigma \mathrm{d}^{d-1}x\sqrt{h}(1+\alpha R)
+\frac{\alpha}{2 G}\int_{\p \Sigma}K,
\ee
where $K$ is the trace of the extrinsic curvature of $\Sigma$ and $R$ is intrinsic curvature of $
\Sigma$. Note, however, that in previous studies, the dual field theory lives on a flat
$d$-dimensional boundary, and what we would like to study in the following is the case when the
dual theory lives on a $d$-dimensional spacetime with compact spatial section.

To be more specific, we shall restrict ourselves to the spherically symmetric (i.e. $k=1$)
(4+1)-dimensional GB-AdS black hole spacetime with line element
\be \label{gb4}
\mathrm{d}s^2=-f_{(5)}(r)\mathrm{d}t^2+\frac{\mathrm{d}r^2}{f_{(5)}(r)}
+r^2(\mathrm{d}\theta^2+\sin^2\theta(\mathrm{d}\varphi^2+\sin^2\varphi\, \mathrm{d}\omega^2)).
\ee
The subsystem $A$ is a subset on the boundary of the bulk spacetime at $r=r_0$ (here $r_0$
plays the role of UV cutoff) and
is chosen to have a spherical boundary $S^2$ which plays the role of entangling surface.
Therefore, in coordinates as used in \eqref{gb4}, the entangling surface can be
parameterized as a constant $\theta$ hypersurface $\theta=\theta_0$ with coordinates
$0\leq\varphi\leq \pi, \, 0\leq \omega<2\pi$. Let us remark that, in principle, the boundary of
the bulk spacetime should be taken at $r=\infty$. However, setting $r=\infty$ directly in the
metric would effectively remove the $M$ dependence and meanwhile render most of the metric
components divergent. Therefore taking an appropriate UV cutoff at $r=r_0$ is a usual practice.

Because of the spherical symmetries, the radial coordinates at any point on $\Sigma$
depends only on $\theta$ but not on $\varphi$ and $\omega$. Therefore, the induced metric on
$\Sigma$ can be written as
\be
h_{ab}\mathrm{d}x^a\mathrm{d}x^b=\bigg(\frac{1}{f_{(5)}(r(\theta))}
r^{\prime 2}(\theta)+r^2(\theta)\bigg)
\mathrm{d}\theta^2+r(\theta)^2\sin^2\theta \mathrm{d}\Omega^2,
\ee
where the prime denotes the derivative with respect to $\theta$ and $\mathrm{d}\Omega^2$ is
the line element on a unit two-sphere.
The scalar curvature of $\Sigma$ can be calculated as
\be
R=2e^{-2F}-4\nabla^2F-6(\p F)^2,
\ee
where
\be
e^{2F}=r^2(\theta) \sin^2\theta,~~\nabla^2F=\frac{1}{\sqrt{h_{\theta\theta}}}
\p_\theta(\sqrt{h_{\theta\theta}}h^{\theta\theta}\p_\theta F),
~~(\p F)^2=h^{\theta\theta}(\p_\theta F)^2.
\ee
To obtain the extrinsic curvature $K$, we define the outward pointing normal vector at
$\p \Sigma$:
\be
n_a=\sqrt{h_{\theta\theta}}\delta_{\theta a},~~ n^a=\sqrt{h^{\theta\theta}}\delta_{\theta a}.
\ee
Using this normal vector, the extrinsic curvature $K$ on $\p \Sigma$ is defined as
\bea
K=(h^{ab}-n^an^b)\nabla_a n_b.
\eea

Combining the above data, the HEE formula Eq.(\ref{GBHEE}) for the subsystem $A$ can
be rearranged into
\bea
&&S_{A}={\pi}\int_0^{\theta_0}\mathrm{d}\theta
\bigg[\left(\frac{r^{\prime 2}}{f_{(5)}}+r^2\right)^{1/2}(2\alpha+r^2\sin^2\theta)\nonumber\\
&&\qquad\qquad\qquad\qquad +2\alpha
\left(\frac{r^{\prime 2}}{f_{(5)}}+r^2\right)^{-1/2}\left(r\cos\theta+r^\prime\sin\theta
\right)^2\bigg], \label{sphee}
\eea
which is to be minimized and integrated out. Notice that we have set the Newton constant $G=1$
in the last formula.

The minimization of $S_A$ involves a variational process which yields a very complicated
second order differential equation for the function $r(\theta)$. This differential equation
should then be solved using the boundary conditions
\be
r^\prime(0)=0,~~~r(\theta_0)=r_0
\ee
and then be substituted back into \eqref{sphee} to get the final result. However,
as $r_0$ approaches infinity, the direct evaluation of $S_A$ will become divergence.
Thus a regularization by subtracting the entanglement entropy corresponding to the
``zero mass black hole'' (i.e. pure AdS background) is necessary. The outcome of the combined
operations as described above will be the regularized HEE
\be
\delta S=S_A-S_A|_{M=0}.
\ee
Due to the overwhelming complicatedness of the related differential equation, the only
way to work out the above process is to resort to numerical methods.

\section{Numerical results}

In this section, we shall study the temperature vs regularized HEE relationship using numerical
method. For RN-AdS \cite{Nguyen:2015wfa} and massive gravity \cite{Zeng:2015tfj},
the $T-\delta S$ curve was shown to possess van der Waals-like behavior at $T<T_c$, and
it was conjectured that an equal area law might also hold on the $T-\delta S$ plane because
numerical calculations show that the relative disagreement between the area of the two
closed regions formed by the $T-\delta S$ curve and the $T=T^\ast$ line is as small as 1\%,
where $T^\ast$ is the same phase equilibrium temperature in the corresponding extended phase
space thermodynamics of the black hole spacetime. Now we would like to see whether similar
van der Waals behavior and/or the equal areal law appears in the situation of spherical
GB-AdS black hole spacetime.

We shall still focus on the $(4+1)$-dimensional case. Within the parameter region
$0.7 < T/T_c < 1.3, \, 0.7< P/P_c<1.3$, the numerical value of the radius $r_+$ of the event
horizon of the spherical GB-AdS black hole can be shown to be less than 10. Thus we set
$r_0=300$, which is large enough as compared to the radius of the event horizon.

\begin{figure}[htbp]
\centering
\includegraphics[width=\textwidth]{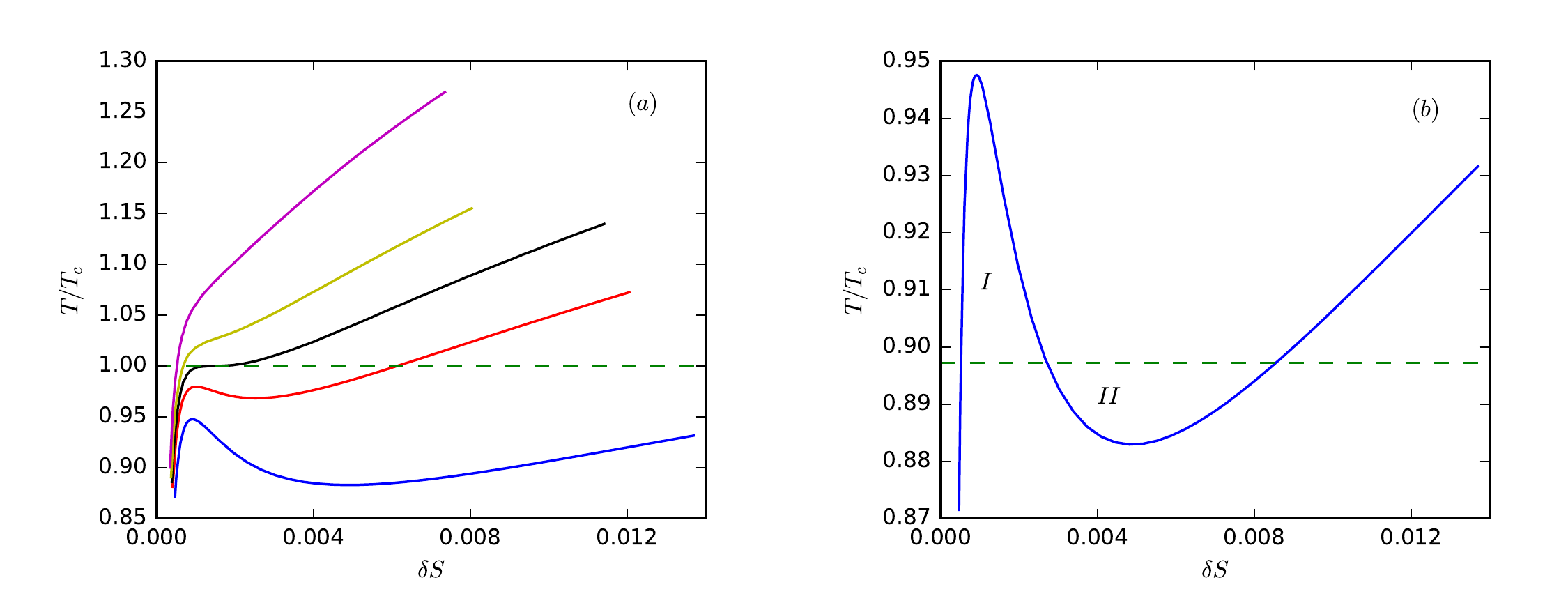}
\caption{$T-\delta S$ curves at $\theta_0=0.1$. (a) From bottom to top, the pressures
corresponding to each curves are:
$0.7P_c$, $0.9P_c$, $P_c$, $1.1P_c$ and $1.3P_c$. The   dashed horizontal line is located
at $T=T_c$. (b) Zoom in of the $P=0.7P_c$ curve in (a). The dashed horizontal line
corresponds to $T=T^*$. }\label{fig:11}
\end{figure}

\begin{figure}[!hbp]
\centering
\includegraphics[width=\textwidth]{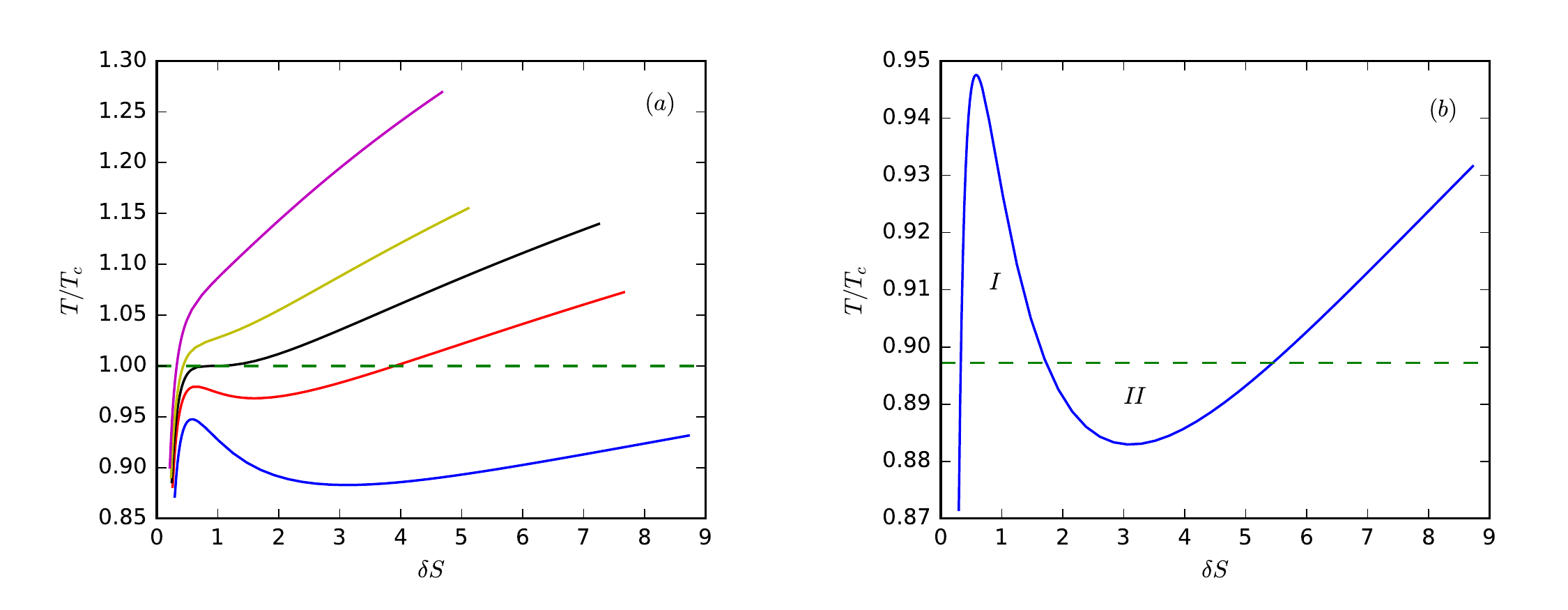}
\caption{$T-\delta S$ plots at $\theta_0=0.5$. (a) From bottom to top, the pressures are: 
$0.7P_c
$, $0.9P_c$, $P_c$, $1.1P_c$ and $1.3P_c$. The dashed horizontal
line corresponds to $T=T_c$. (b) Zoom in of the $P=0.7P_c$ curve in (a). The dashed horizontal
line corresponds to $T=T^*$.}\label{fig:13}
\end{figure}

In Fig.\ref{fig:11} and Fig.\ref{fig:13}, we present the plots of temperature versus
regularized HEE $\delta S$ at several fixed pressures. These two figures corresponds to
different sizes of the entanglement surface characterized by $\theta_0=0.1$ and
$\theta_0=0.5$ respectively. It can be seen that,
when the pressure is lower than the critical pressure $P_c$, there can be three different
$\delta S$ if the temperature takes value in a certain range. As $P$ approaches $P_c$ from
below, the three different values of $\delta S$ move closer, until they merge into a single
value at $P=P_c$. When $P>P_c$, $T$ becomes monotonic in $\delta S$. Such behavior is
qualitatively similar to the temperature versus black hole entropy curves as shown in
Fig.\ref{fig:1}.

Next we would like to examine whether there is an equal area law on the $T-\delta S$ plane.
The two closed regions I and II formed by the $T-\delta S$ curve and the horizontal
line $T=T^\ast$ are explicitly marked in Fig.\ref{fig:11}(b) and Fig.\ref{fig:13}(b) for
the cases $\theta_0=0.1$ and $\theta_0=0.5$ respectively. The phase 
equilibrium temperature $T^\ast$,
the areas of both closed regions and their relative disagreement are presented numerically
at the given pressures $P/P_c=0.6,\, 0.7, \, 0.8,\, 0.9$ respectively in Table \ref{tab1} and
Table \ref{tab3}. It can be seen from these tables that, as the pressure approaches $P_c$
from below, the relative disagreement between the areas of the two closed regions
decreases. However, at lower pressures, the relative disagreement can become significantly 
large, and consequently the equal area law cannot hold on the $T-\delta S$ plane.

\begin{table}[hbp]
\centering
\begin{tabular}{ccccc}
\hline \hline
$P/P_c$&$T^*/T_c$ & A(I) & A(II) & (A(I)-A(II))/A(I) \\
\hline
0.6&0.8485& $1.2490\times10^{-4}$ &$1.1468\times10^{-4}$ &8.18\%\\
0.7&0.8972& $5.7431\times10^{-5}$ &$5.4452\times10^{-5}$ &5.12\%\\
0.8&0.9381& $2.1475\times10^{-5}$ &$2.0790\times10^{-5}$ &3.19\%\\
0.9&0.9721& $4.5825\times10^{-6}$ &$4.5162\times10^{-6}$ &1.45\%\\
\hline\hline
\end{tabular}
\caption{Areas and their relative disagreement for the regions I and II at
$\theta_0=0.1$. }\label{tab1}
\end{table}

\begin{table}[!hbp]
\centering
\begin{tabular}{ccccc}
\hline \hline
$P/P_c$&$T^*/T_c$ & A(I) & A(II) & (A(I)-A(II))/A(I) \\
\hline
0.6&0.8485& $7.9775\times10^{-2}$ &$7.3208\times10^{-2}$ &8.25\%\\
0.7&0.8972& $3.6551\times10^{-2}$ &$3.4670\times10^{-2}$ &5.14\%\\
0.8&0.9381& $1.3582\times10^{-2}$ &$1.3229\times10^{-2}$ &2.59\%\\
0.9&0.9721& $2.9091\times10^{-3}$ &$2.8782\times10^{-3}$ &1.06\%\\
\hline\hline
\end{tabular}
\caption{Areas and their relative disagreement for the regions I and II at
$\theta_0=0.5$. }\label{tab3}
\end{table}

Although we presented the numerical results only for two distinct values of $\theta_0=0.1,\,0.5$,
this does not imply that these values of $\theta_0$ are special in any sense. Actually we have
carried out the numerical process for some other values of $\theta_0$, and the results are
qualitatively the same. Therefore we conclude that the break down of equal area law
on the $T-\delta S$ plane should be a generic phenomenon for the HEE associated with the
spherically symmetric GB-AdS black hole spacetime.

Besides the qualitative $T-\delta S$ behavior, let us examine another important aspect of
HEE in the case of spherically symmetric GB-AdS black holes, i.e. the so-called entanglement
thermodynamics.   In Ref. \cite{Bhattacharya:2012mi} (see also \cite{Guo:2013aca}), the first law
of entanglement entropy has been proposed, which states that the increase of HEE is proportional
to the increase of the energy of the subsystem, i.e. $\delta S\propto \Delta  E_A$, provided
$ml^d\ll 1$, where $m$ is proportional to the black hole mass and $l$ is related to
the size of the
subsystem $A$ which is of the order $z=z_*$ in $d=4$ ($z_*$ is the Poincare radial coordinate
which marks the position of the co-dimension 2 hypersurface). In our convention, we have
$r\sim 1/z$ and so  the above condition becomes $\frac{M}{r(0)^d}\ll 1 $ (we still take $d=4$).
It should be noted that the calculation in \cite{Bhattacharya:2012mi} was performed in spacetime
with planar boundary in contrast to the compact boundary in our case. Therefore,
it is interesting to see whether the first law of entanglement entropy still holds
in our case. It should be remarked that the increase $\Delta E_A$ of the subsystem $A$ of the
dual field theory is proportional to the black hole mass $M$ \cite{Guo:2013aca,Balasubramanian:1999re},
therefore what we would like to explore is whether there is a linear relationship between the
regularized HEE $\delta S$ and the black hole mass $M$.

\begin{figure}[htbp]
\centering
\includegraphics[width=\textwidth]{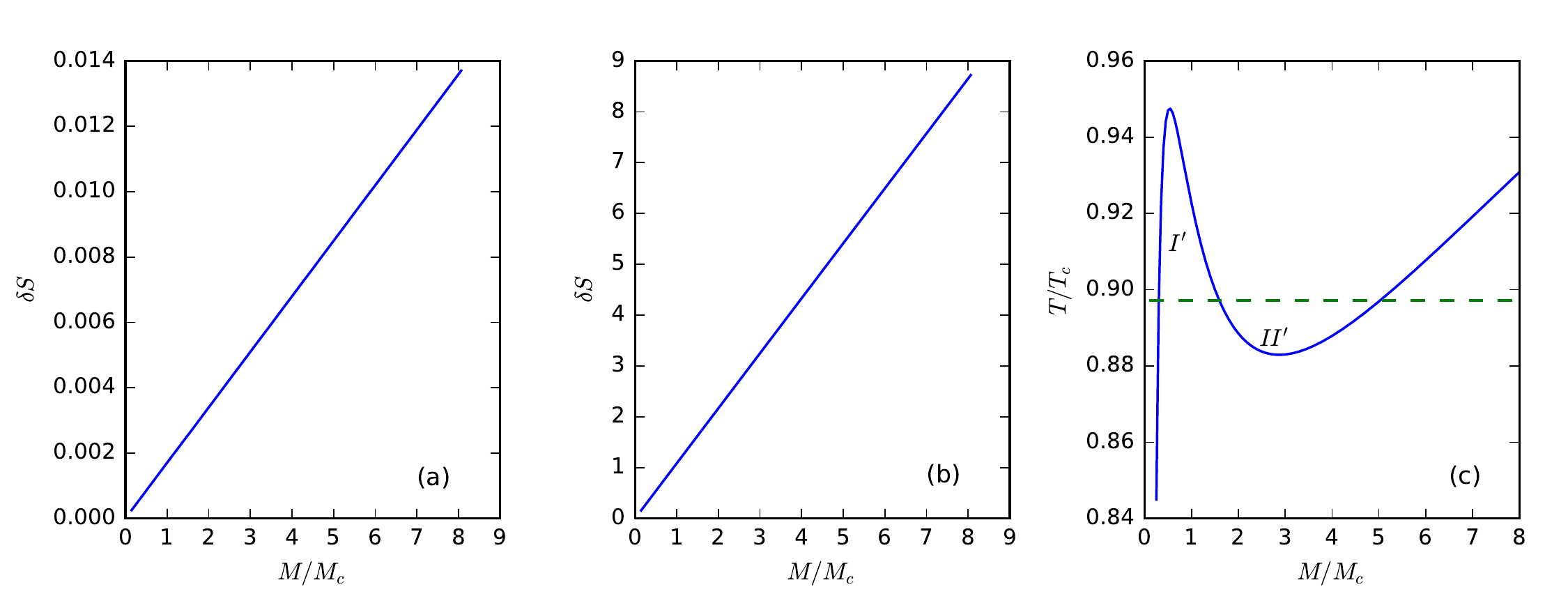}
\caption{$\delta S$ and $T$ versus $M$ at $P=0.7P_c$. (a) $\delta S$ vs $M$ with $\theta_0=0.1$;
(b) $\delta S$ vs $M$ with $\theta_0=0.5$. (c) $T$ vs $M$, the dashed
horizontal line corresponds to $T=T^*\simeq 0.8972\,T_c$. The two closed regions are
marked as I$^{\prime}$ and II$^\prime$ respectively.}\label{fig:linear}
\end{figure}

Fig.\ref{fig:linear}(a) and Fig. \ref{fig:linear}(b) give the plot of $\delta S$ as
function of $M$ at the fixed pressure $P/P_c=0.7$ in the cases of $\theta_0=0.1$ and
$\theta_0=0.5$ respectively. It turns out that for both small and large subsystem $A$,
there is a very good linear relationship between $\delta S$ and the black hole mass $M$.
In fact, this linear relationship shows up under all parameter ranges as presented in Tables
\ref{tab1} and \ref{tab3}. This indicates that the first law of entanglement entropy indeed
holds in our case.

\begin{table}[!hbp]
\centering
\begin{tabular}{ccccc}
\hline \hline
$P/P_c$&$T^*/T_c$ & A(I$^\prime$) & A(II$^\prime$) & (A(I$^\prime$)-A(II$^\prime$))/A(I$^\prime$) \\
\hline
0.6&0.8485& $6.8084\times10^{-2}$ &$6.2521\times10^{-2}$ &8.17\%\\
0.7&0.8972& $3.3759\times10^{-2}$ &$3.2034\times10^{-2}$ &5.11\%\\
0.8&0.9381& $1.3451\times10^{-2}$ &$1.3090\times10^{-2}$ &2.68\%\\
0.9&0.9721& $3.0562\times10^{-3}$ &$3.0282\times10^{-3}$ &0.91\%\\
\hline\hline
\end{tabular}
\caption{Areas and their relative disagreement for the regions I$^\prime$ and II$^\prime$
on $T-M$ plane.}
\label{tab4}
\end{table}

Now since $\delta S $ is  proportional to $M$, one naturally expects that
the $T-M$ relationship must be qualitatively similar to
the $T-\delta S$ relationship. A little calculation indicates that
\bea
&&\bigg(\frac{\partial T}{\partial M}\bigg)_{P}
=\bigg(\frac{\partial T}{\partial S}\bigg)_{P}\bigg(\frac{\partial S}{\partial M}\bigg)_{P}=0,\\
&&\bigg(\frac{\partial^2 T}{\partial M^2}\bigg)_{P}
=\bigg(\frac{\partial T}{\partial S}\bigg)_{P}\bigg(\frac{\partial^2 S}{\partial M^2}\bigg)_{P}
+\bigg(\frac{\partial^2 T}{\partial S^2}\bigg)_{P}
\bigg(\frac{\partial S}{\partial M}\bigg)_{P}^2=0,
\eea
where the last equality holds at the critical point on the $T-S$ plane, thanks to
Eq.(\ref{cri}). Exploring the expressions (\ref{mass}) and (\ref{T1}), one can show
that there is an oscillatory segment on each $T-M$ curve if $P<P_c$. Furthermore,
comparing the areas of the closed regions I$^\prime$ and region II$^\prime$ in
in Fig.\ref{fig:linear}(c), we find that the relative disagreements are roughly of the same
order as in the cases of the $T-\delta S$ curves. For detailed numerical results, see
Table \ref{tab4}.

It should be pointed out that there is no reason to think of the areas
A(I$^\prime$) and A(II$^\prime$) as being equal in black hole thermodynamics.
Therefore, the linear relationship between $\delta S$ and $M$ may serve as a good reason
in judging that there is no equal area law on the $T-\delta S$ plots.

At this stage, it seems necessary to address the following problem: why the equal area law
on the $T-\delta S$ plane seems to hold in the case of charged AdS black hole
\cite{Nguyen:2015wfa} whilst it is not the case for spherically symmetric Gauss-Bonnet black
hole? To our understanding, this apparent controversy has nothing to do with the choice
of gravity models. The seemingly holding equal area law in \cite{Nguyen:2015wfa} is
the consequence of the insufficient exploration on the range of the charge parameter $Q$
(which plays similar role as the pressure $P$ in the present work). In Fig.2 of
\cite{Nguyen:2015wfa}, only the case $Q=0.9Q_c$ had been checked numerically (however with
several values of $\theta_0$). Had we explored
only the case $P=0.9P_c$ in our work, one might draw the conclusion that the equal area law
should also hold on the $T-\delta S$ plane for spherically symmetric Gauss-Bonnet black
hole up to the relative disagreement around $1\%$ (see the bottom rows of Tables \ref{tab1}
and \ref{tab3}). In fact, if one looks only at these bottom rows, the relative disagreements
between the two closed areas are even smaller than that found in \cite{Nguyen:2015wfa}.
However, exploring broader ranges of the
parameter $P$ has revealed the fact that the equal area law does not hold actually.

\begin{table}[!hbp]
\centering
\begin{tabular}{ccccc}
\hline \hline
$Q/Q_c$&$T^*/T_c$ & A(I) & A(II) & (A(I)-A(II))/A(I) \\
\hline
0.6&1.0955& $3.0415\times10^{-5}$ &$2.5353\times10^{-5}$ &16.64\%\\
0.7&1.0724& $1.5835\times10^{-5}$ &$1.4180\times10^{-5}$ &10.45\%\\
0.8&1.0488& $6.6235\times10^{-6}$ &$6.2605\times10^{-6}$ &5.48\%\\
0.9&1.0247& $1.5817\times10^{-6}$ &$1.5528\times10^{-6}$ &1.86\%\\
\hline\hline
\end{tabular} 
\caption{A check for the ``equal area law'' on the $T-\delta S$ plane for (3+1)D RN-AdS 
black hole with $\theta_0=0.1$ and $\theta_c=0.099$.  $Q_c$ and $T_c$ are the critical 
charge and critical temperature respectively whose value are 
presented in \cite{Nguyen:2015wfa}. $T^*$ is the RN-AdS black hole phase equilibrium  
temperature. } \label{rn1}
\end{table}

\begin{table}[!htbp]
\centering
\begin{tabular}{ccccc}
\hline \hline
$Q/Q_c$&$T^*/T_c$ & A(I$^\prime$) & A(II$^\prime$) & (A(I$^\prime$)-A(II$^\prime$))/A(I$^\prime$) \\
\hline
0.6&1.09545& $3.9449\times10^{-3}$ &$3.2885\times10^{-3}$ &16.64\%\\
0.7&1.07238& $2.0538\times10^{-3}$ &$1.8393\times10^{-3}$ &10.44\%\\
0.8&1.04881& $8.5909\times10^{-4}$ &$8.1203\times10^{-4}$ &5.48\%\\
0.9&1.02470& $2.0515\times10^{-4}$ &$2.0134\times10^{-4}$ &1.86\%\\
\hline\hline
\end{tabular} 
\caption{The case for RN-AdS black hole on the $T-M$ plane. }\label{rn2}
\end{table}

To support the above statements, we have re-worked out the numerics for the RN-AdS black hole.
One thing to remark here is that 
\cite{Nguyen:2015wfa} has used a different cutoff scheme, i.e. instead of introducing a 
UV cutoff over $r$ as we do in this paper, the author of \cite{Nguyen:2015wfa} has chosen to
use a cutoff $\theta_c$ over the angular variable $\theta$. Anyway, complying completely with 
the conventions of \cite{Nguyen:2015wfa}, we obtained numerical results which fully support
the statements given in the last paragraph. Table \ref{rn1} contains the numerical results for 
the two closed areas on the $T-\delta S$ plane and their relative disagreements 
for (3+1)D RN-AdS black hole. For simplicity, we present the results only for $\theta_0=0.1$ 
and $\theta_c=0.099$, the cases for other choices of $\theta_0$ have also been checked 
and the results are qualitatively the same. The value of the charge parameter varies 
from $0.6\,Q_c$ to $0.9\,Q_c$, where the bottom row corresponds exactly to the same parameter 
set presented in the first row of Table 1 of \cite{Nguyen:2015wfa} (A(I) and A(II) respectively 
are denoted $A_1$ and $A_2$ in \cite{Nguyen:2015wfa})\footnote{The numerical values for 
A(I) and A(II) in the bottom row of Table \ref{rn1} are about 0.6\% less than the values
given in \cite{Nguyen:2015wfa}. We suspect that this difference might be originated 
from the choice of different numerical integration algorithms. 
The numerical program which we use for producing the table is available upon request, 
just send us an e-mail.}. Table \ref{rn2} presents the parallel 
results on the $T-M$ plane. From the last two tables one can see that, if broader ranges for 
the parameter $Q$ had been explored in \cite{Nguyen:2015wfa}, one would not have drawn 
the conclusion that the equal area law holds on the $T-\delta S$ plane (nor on the $T-M$ plane).
The reason for the breakdown for the equal area law in the RN-AdS case can also be attributed to
the linear relationship between $\delta S$ and $M$, which has also been numerically
checked to hold at very high accuracy.

\section{Concluding remarks}

In this work, we extended the study of analogy between black hole entropy and HEE
to (4+1)-dimensional spherically symmetric Gauss-Bonnet AdS black hole spacetime.
The thermodynamics of the black hole is reviewed, which emphasis on the behaviors on
the $T-S$ plane. Then the regularized HEE $\delta S$ against the black hole temperature
is calculated numerically. The results show that the isobaric $T-\delta S$ curves behave
qualitatively the same as the isobaric $T-S$ curve and exhibits van der Waals-like structure,
however there is no reason to believe there is an analogy of equal area law on the
$T-\delta S$ plane.

We also find that the regularized HEE $\delta S$ is proportional to the black hole mass $M$,
which may be understood as the first law of holographic entanglement entropy. Note that
for a spacetime with compact boundary, the first law of holographic entanglement entropy
has never been established before. The linear relation between $\delta S$ and $M$ 
may also serve as an
explanation for the failure of equal area law on the $T-\delta S$ plane.

\section*{Notes added}

After the first version of this manuscript has appeared on arXiv, we have noticed the more 
recent work \cite{Zeng:2016aly}, which also studied the holographic entanglement entropy versus
temperature relationship for the (charged) spherically symmetric Gauss-Bonnet black holes. 
There the authors claimed the correctness of the equal area law on the $T-\delta S$ plane.
However, the areas compared in that paper are $A_L$ and $A_R$ defined in \eqref{alr}
(with $S$ replaced by $\delta S$), rather than the correct A(I) and A(II) used in the present 
paper. The ``relative error'' between $A_L$ and $A_R$ differs from the relative disagreement
between A(I) and A(II) by adding the whole area below the isotherm $T=T^\ast$ (which is a 
huge quantity as compared to the area of the closed region) in the denominator, this 
explains why the relative errors presented in \cite{Zeng:2016aly} are so small.

\acknowledgments

We would like to Xiao-Xiong Zeng for helpful discussions on the numerical methods.
The numerical calculations in this work are carried out using both Mathematica and the
Anaconda Scientific Python Distribution, which yield perfect agreements.
This work is supported by the National Natural Science Foundation of China under the grant
No. 11575088.


\end{document}